\documentclass[11pt]{article}
\usepackage{epsfig}

\newcommand{\BABARPubYear}    {06}

\newcommand{\BABARConfNumber} {024}
\newcommand{\SLACPubNumber} {12004}

\input babarsym

\long\def\inst#1{\par\nobreak\kern 4pt\nobreak
    {\it #1}\par\vskip 10pt plus 3pt minus 3pt}

\begin{document}
{\pagestyle{empty}

\begin{flushright}
\babar-CONF-\BABARPubYear/\BABARConfNumber \\
SLAC-PUB-\SLACPubNumber \\
\end{flushright}

\par\vskip 5cm

\begin{center}

\Large \bf  Measurement of the Relative Branching Fractions for $B^- \rightarrow
 D/D^{*}/D^{**}(D^{(*)}\pi) \ell^- \bar{\nu}_{\ell}$  with a Large Sample of Tagged B Mesons
\end{center}
\bigskip

\begin{center}
\large The \babar\ Collaboration\\
\mbox{ }\\
\today
\end{center}
\bigskip \bigskip

\begin{center}
\large \bf Abstract
\end{center}

We present a study of B semileptonic decays into charm final states based on 
211.7 fb$^{-1}$ of data collected at the $\Upsilon(4S)$ resonance with the \babar\
detector at the \pep2\ $e^+e^-$ storage ring. 
Using a novel technique 
 based on the simultaneous fit of a set of variables reconstructed on the 
 recoil of a $B$ tagged in an hadronic decay mode, we measure the 
 relative branching fractions $\Gamma(B^- \rightarrow
 D^0\ell^- \bar{\nu}_{\ell})/\Gamma(B^- \rightarrow D X \ell^- \bar{\nu}_{\ell}) = 0.210 \pm 0.017~(\mbox{stat.})  \pm  0.021~(\mbox{syst.}) $, $\Gamma(B^- \rightarrow
 D^{*0}\ell^- \bar{\nu}_{\ell})/\Gamma(B^- \rightarrow D X \ell^- \bar{\nu}_{\ell}) = 0.611 \pm  0.022~(\mbox{stat.}) \pm 0.027~(\mbox{syst.}) $ and $\Gamma(B^- \rightarrow
D^{**0}(D^{(*)}\pi)\ell^- \bar{\nu}_{\ell})/\Gamma(B^- \rightarrow D X \ell^- \bar{\nu}_{\ell}) = 0.173 \pm 0.017~ (\mbox{stat.}) \pm 0.021~(\mbox{syst.}) $. 
\vfill
\begin{center}

Submitted to the 33$^{\rm rd}$ International Conference on High-Energy Physics, ICHEP 06,\\
26 July---2 August 2006, Moscow, Russia.

\end{center}

\vspace{1.0cm}
\begin{center}
{\em Stanford Linear Accelerator Center, Stanford University, 
Stanford, CA 94309} \\ \vspace{0.1cm}\hrule\vspace{0.1cm}
Work supported in part by Department of Energy contract DE-AC03-76SF00515.
\end{center}

\newpage
} 

\begin{center}
\small

The \babar\ Collaboration,
\bigskip

%
{B.~Aubert,}
{R.~Barate,}
{M.~Bona,}
{D.~Boutigny,}
{F.~Couderc,}
{Y.~Karyotakis,}
{J.~P.~Lees,}
{V.~Poireau,}
{V.~Tisserand,}
{A.~Zghiche}
\inst{Laboratoire de Physique des Particules, IN2P3/CNRS et Universit\'e de Savoie,
 F-74941 Annecy-Le-Vieux, France }
% ...

\end{center}\newpage

\section{Introduction}
\label{sec:Introduction}

The determination of the individual exclusive branching fractions of
$B \to X_c \ell \bar{\nu}_{\ell}$ decays\footnote{Here $X_c$ refers to any charm state, $X_u$ to any charmless particle.} is important for the study of the 
dynamics of semileptonic decays of the $B$ meson. Precise data on 
the spectroscopy of the hadronic system is needed to be compared with 
theoretical predictions, obtained under different 
assumptions~\cite{Morenas:1997nk,Leibovich:1997em, Ebert:2000bj,DiPierro:2002eu}.
This is especially important since the mass of
the hadronic system, recoiling against the leptonic pair in the decay,
is a crucial parameter in the extraction of $|V_{cb}|$ in exclusive
semileptonic decays, in isolating $B \to X_u \ell \bar{\nu}_{\ell}$ from
$B \to X_c \ell \bar{\nu}_{\ell}$ decays for determining $|V_{ub}|$, and also in
the extraction of the heavy quark masses and other non-perturbative Operator Product Expansion 
parameters from the distribution of spectral moments in semileptonic
decays. This is exemplified by the fact that one of the leading sources of
systematic uncertainty in the extraction of $|V_{cb}|$ from the exclusive
decay $\bar B \to D^* \ell \bar{\nu}_{\ell}$ is our limited knowledge of the
background due to $\bar B \to D^* \pi \ell \bar{\nu}_{\ell}$~\cite{pdg}. Reducing 
this uncertainty is of crucial importance, as lattice calculations
promise to improve the theoretical accuracy on the form factor normalization to
2-3~\%. Improvements in our knowledge of $B \to X_c \ell \bar{\nu}_{\ell}$ decays will
also benefit the accuracy in the extraction of $|V_{ub}|$, as analyses are
extending their probe into kinematical regions where these decays represent a sizable background.

The first determination of the fractions of $D^{**}$ states in semileptonic
$b$ decays was obtained by the ALEPH~\cite{Buskulic:1996uk},
DELPHI~\cite{Abreu:2000,Bloch:2000} and OPAL~\cite{Abbiendi:2002ge}
experiments at LEP, where some information on the mass distributions of these
states was also extracted, and by CLEO~\cite{Anastassov:1997im} at CESR.
In particular, the kinematics at LEP, where the $B$ hadrons
had a significant boost and were separated in opposite hemispheres, allowed
analyses based on the topology of the particle tracks with respect to the decay
vertices. On the other hand, primary fragmentation particles represented a
background and statistics were limited. More recently, new 
results have been obtained by the D0~\cite{Abazov:2005ga}
experiment at the Tevatron.

Not only are the data samples at the $B$ factories much larger than those
obtained at earlier collider experiments, in addition the feasibility to fully
reconstruct one of the two $B$ mesons produced exclusively in the decay of the $\Upsilon(4S)$ resonance permits 
the study of semileptonic decays of the other $B$ meson in an almost unbiased way.
This allows the study of exclusive semileptonic $B$ decays involving charm mesons, specifically $D$, $D^*$,  
resonant $D^{**}$ and non-resonant $D^{(*)} \pi$ with the \babar\ data.

Measurements of the largest $B$ meson branching fraction, $\mathcal{B}$($\bar B^0 \to D^{*-} \ell^+ \nu_{\ell}$), by the CLEO, the LEP and the \babar\ and BELLE experiments, based  on different reconstruction techniques, need to be 
reconciled. The current average value~\cite{hfag} has a $\chi^2$/ndof = 14.8/7 with a probability of
just 3.8~\%. 

Furthermore, there have been so far only very few
studies of semileptonic decays to $D^{**}$ and non-resonant final states at the 
$B$-factories. BELLE has recently reported the determination of the
$D^{(*)} \pi \ell \bar{\nu}_{\ell}$ decay branching fraction~\cite{Abe:2005up}.

This study uses a novel technique to extract the exclusive branching
fractions\footnote{Charge-conjugate modes are implied throughout this paper, unless explicitly stated otherwise.} for $B^- \to D^0 \ell^- \bar{\nu}_{\ell}$, $B^- \to D^{*0} \ell^- \bar{\nu}_{\ell}$  and
$B^- \to D^{**0} \ell^- \bar{\nu}_{\ell}$ ($\ell = e$ or $\mu$), where $D^{**0}$ denotes here an hadronic final
state containing a charm meson and with total mass above that of the $D^*$
state, therefore including both  $D^{**0}$ mesons and non-resonant
$D^{(*)} \pi$ states. 

\section{The \babar\ Detector and Dataset}
\label{sec:babar}
This analysis is based on data collected with the \babar\ detector at the \pep2\ storage ring. The total integrated luminosity of the data set is 211.7 fb$^{-1}$ collected on the $\Upsilon(4S)$. The corresponding number of produced $B\bar{B}$ pairs is 239 million. The \babar\ detector is described in detail elsewhere~\cite{ref:babar}. Charged-particle trajectories are measured by a 5-layer double-sided silicon vertex tracker (SVT) and a 40-layer drift chamber (DCH), both operating in a 1.5-T solenoidal magnetic field. Charged-particle identification is provided by the average energy loss (d$E$/d$x$) in the tracking devices and by an internally reflecting ring-imaging Cherenkov detector. Photons are detected by a CsI(Tl) electromagnetic calorimeter. Muons are identified by the instrumented magnetic-flux return (IFR). 
We use 
Monte Carlo simulations of the \babar\ detector based on {\tt GEANT}~\cite{geant} to validate the event reconstruction and the fitting technique used to extract the individual semileptonic branching fractions. 

\section{Overview of the Analysis Method}
\label{sec:Analysis}

The technique developed for this analysis identifies a set of variables, which can be used to discriminate between the individual semileptonic decay modes. 
These variables are 
 i) the missing mass squared reconstructed with respect to the $D \ell$
system, $m^2_{miss}$, ii) the lepton momentum in the $B$ rest frame
 and iii) the number of reconstructed tracks in addition to 
those used for reconstructing the $D$ state and the lepton.
In order to reduce the
sensitivity to Monte Carlo modelling, shapes of the discriminating
variables for the $D^0$, $D^{*0}$ and $D^{**0}$ decays are extracted from data samples that are highly enriched with the individual decay modes.

Semileptonic $B$ decays containing
one fully reconstructed $D$ meson and recoiling against a fully reconstructed
$B$ in hadronic decay modes ($B_{tag}$) are selected. We reconstruct $B_{tag}$ decays of the type $B \rightarrow \bar{D} Y$, where $D$ refers to a charm meson, and $Y$ represents a collection of hadrons with a total charge of $\pm 1$, composed of $n_1\pi^{\pm}+n_2 K^{\pm}+n_3 K^0_S+n_4\pi^0$, where $n_1+n_2< 6$, $n_3<3$, and $n_4<3$. Using $D^0$ and $D^{*0}$ as seeds for $B^-$ decays, we reconstruct about 1000 different decay chains. The kinematic consistency of a $B_{tag}$ candidate with a $B$ meson decay is checked using two variables, the beam-energy substituted mass $m_{ES}=\sqrt{s/4-\vec{p}_B^2}$ and the energy difference, $\Delta E = E_B -\sqrt{s}/2$. Here $\sqrt{s}$ refers to the total energy, and $\vec{p}_B$ and $E_B$ denote the momentum and energy of the $B_{tag}$ candidate in the $\Upsilon(4S)$ center of mass frame. For correctly identified $B_{tag}$, the $m_{ES}$ distribution peaks at the $B$ meson mass, while $\Delta E$ is consistent with zero. 
The signal region for tagging $B$ candidates is defined as $m_{ES} > 5.27$ GeV/$c^2$.   

The analysis exploits the presence of two charmed mesons 
in the final state: one used as a seed for the exclusive reconstruction of the $B_{tag}$, 
and another in the $B$ semileptonic decay, to obtain a high reconstruction efficiency. 
The reconstruction starts from the $B$ semileptonic decay, identifying a charm meson of the correct 
charge-flavor correlation with a lepton with momentum in the $\Upsilon(4S)$ frame higher than 0.6 GeV/$c$ and then selects the $B_{tag}$ with the smallest $\Delta E$ among those which do not overlap with the reconstructed charm meson associated with the lepton. 
$D^0$ candidates are reconstructed in nine decay modes: $D^0 \rightarrow K^-\pi^+$,  $K^- \pi^+ \pi^0$, 
$K^- \pi^+ \pi^+ \pi^-$, $K^0_S \pi^+ \pi^-$, $K^0_S \pi^+ \pi^- \pi^0$, $K^0_S \pi^0$, $K^+ K^-$, $\pi^+ \pi^-$, $K^0_S K^0_S$. 
$D^+$ candidates are reconstructed in seven decay modes:  $D^+ \rightarrow K^- \pi^+ \pi^+$, $K^- \pi^+ \pi^+ \pi^0$, $K^0_S \pi^+$ , $K^0_S \pi^+ \pi^0$, $K^+ K^- \pi^+$, $K^0_S K^+$, $K^0_S \pi^+ \pi^+ \pi^-$. 

This procedure is 
designed to maximize the efficiency of the $D \ell$ reconstruction and, as a result, the $B_{tag}$ 
sample shows high purity, without need of further cuts.

Then, $B^- \rightarrow D X \ell^- \bar{\nu}_{\ell}$ events, where $X$ can be either nothing or any particle(s) from a charged $B$ semileptonic decay into an higher mass charm state (or a  non resonant state), are identified by relatively loose selection criteria: we require the charm meson ground state invariant masses $M_{D^0}$ and $M_{D^+}$ to be in the range $1.85 < M_{D^0}< 1.88$ GeV/$c^2$ and $1.853 < M_{D^+}< 1.883$ GeV/$c^2$ and the cosine of the angle between the $D$ candidate and the lepton in the $\Upsilon(4S)$ frame to be less than zero, to reduce background from non $B$ semileptonic decays. 
After these loose selection criteria, the sample contains leptons from prompt and cascade $B$ meson decays (i.e. the lepton does not come directly from the $B$) plus various background sources (e.g. photon conversions and $\pi^0$  Dalitz decays, combinatoric \BB\ background) that need to be subtracted. The contamination from cascade $B$ meson decays (about 15\% of the total sample) is removed by using the simulated Monte Carlo distributions for these backgrounds.  These events are   
reweighted to account for possible differences between the branching fractions used in our Monte Carlo simulation and the latest experimental measurements.
The photon conversion and $\pi^0$ Dalitz decay background (less than 2\% of the total electron sample) is removed by using a dedicated algorithm, which performs the reconstruction of vertices between tracks of opposite charge whose invariant mass is compatible with a photon conversion or a $\pi^0$ Dalitz decay. 

\begin{figure}[!t]
\centering
\epsfig{figure=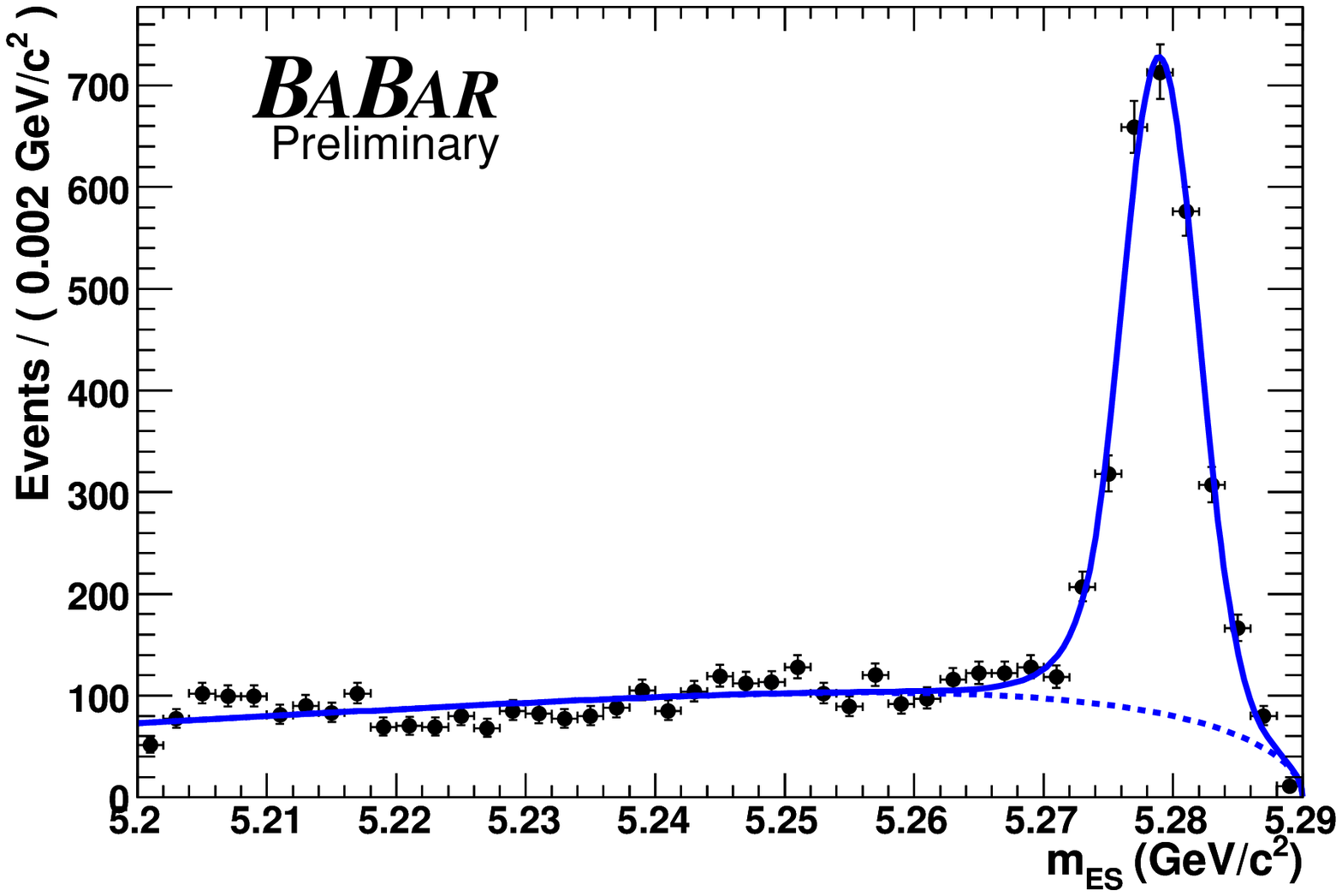,width=7.4cm}
\epsfig{figure=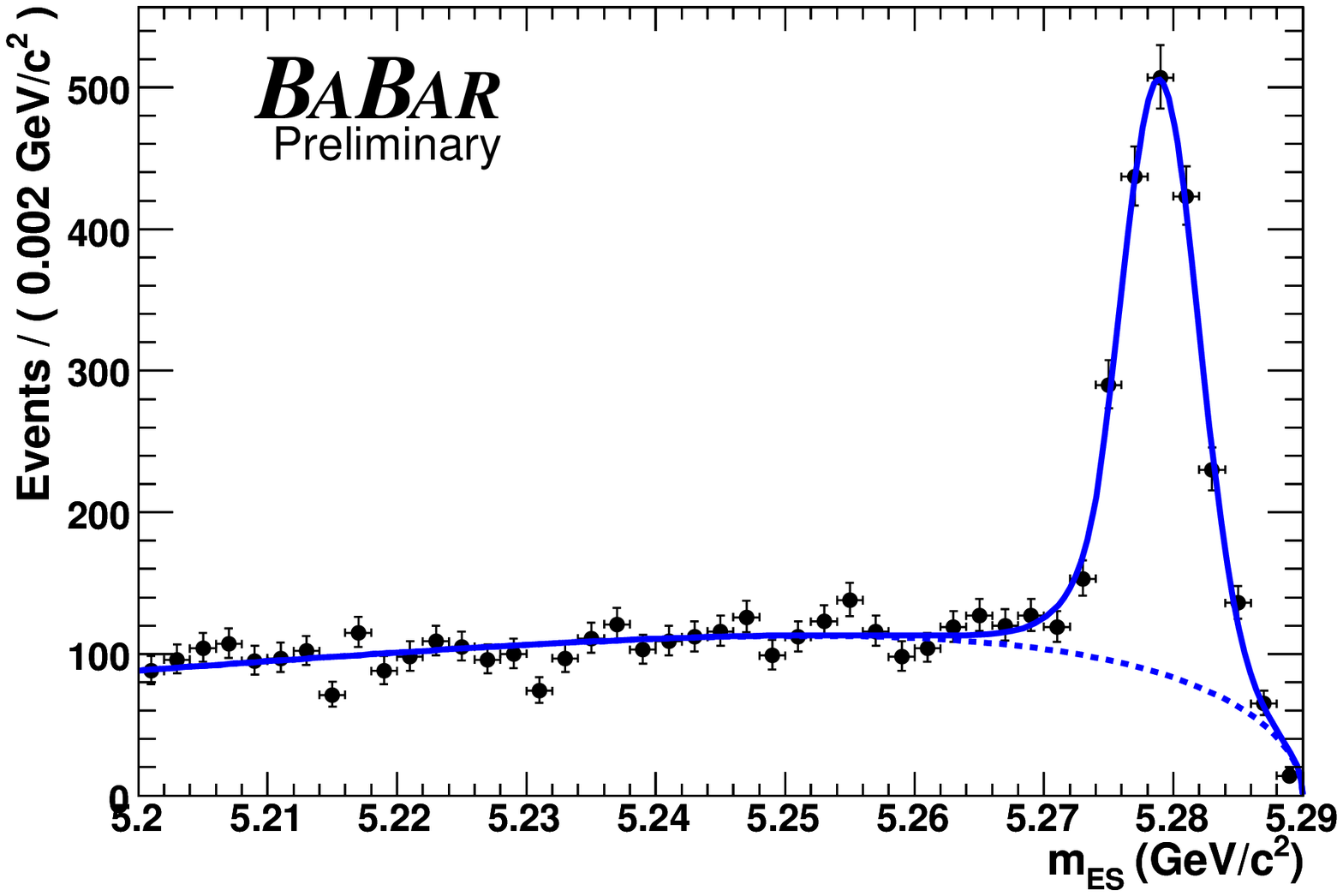,width=7.4cm}
\caption{The $m_{ES}$ distributions (data points with statistical errors) for the hadronic $B_{tag}$ with $B^- \rightarrow D X e^- \bar{\nu}_e$ (left) and $B^- \rightarrow D X \mu^- \bar{\nu}_{\mu}$ (right) events. The solid line represents a fit to the total distribution, the dotted line the combinatorial background. The fitted signal yields are respectively $2409 \pm 122$ and $1624 \pm 130$ for the electron and muon samples.}
\label{fig:mesBplus}
\end{figure}

Figure \ref{fig:mesBplus} shows the $m_{ES}$ distribution for the $B_{tag}$ candidates in the $B^- \rightarrow D X \ell^- \bar{\nu}_{\ell}$ sample. We fit this distribution with empirical functions: a Crystal Ball~\cite{CrystallBall} for the signal and an ARGUS function~\cite{argusfunc} for the background. 
Combinatoric \BB\ background is removed by performing a sideband subtraction by scaling the number of background events in the $B_{tag}$~$m_{ES}$ sideband region $5.21 < m_{ES} < 5.26$ GeV/$c^{2}$ to the integral of the background function in the signal region. 
Cross-feed effects ($B^-_{tag}$ candidates erroneously reconstructed as a neutral $B$) are corrected for using Monte Carlo simulation. We estimate the fraction of $\bar{B}^0$ events in the reconstructed $B^-$ sample to be 7\%.

To determine the total reconstruction efficiencies for individual 
signal decays using Monte Carlo simulation, the fitted yield of events 
in the $m_{ES}$ distribution of the $B_{tag}$ decays is compared to the 
produced number of semileptonic decays in tagged events.

The inclusive distributions of the discriminating variables are determined for the selected $B^- \rightarrow D X \ell^- \bar{\nu}_{\ell}$ samples.
The missing mass squared in the event is defined as:

\begin{equation}
 m^2_{miss} = (p(\Upsilon) -p(B_{tag}) - p(D^{0,+}) - p(\ell))^2
\end{equation}

\noindent in terms of the various particle four-momenta and for correctly measured signal events it corresponds to the invariant mass of the $X \nu_{\ell}$ system in the $B^- \to D^{0,+} X \ell^- \bar{\nu}_{\ell}$ decay. Thus for $B^- \rightarrow D^0 \ell^- \bar{\nu}_{\ell}$ decays, the missing mass squared corresponds to the neutrino mass and  peaks at zero. For semileptonic decays to excited $D$ mesons or non-resonant states, the distribution is shifted towards higher masses. The selection of fully reconstructed
events results in an excellent missing mass resolution of 0.04~GeV$^2$/$c^4$, an order of magnitude lower compared
to non-tagged analyses~\cite{exampArgus}.

The lepton momentum in the $\Upsilon(4S)$ frame is the second variable which can be used to some extent to discriminate among the different $B$ semileptonic decays. Leptons directly produced from $B$ decays will have momenta greater than secondary cascade decays of the type $B \rightarrow D \rightarrow \ell$. For true $B$ semileptonic decays, the lepton momentum is peaked between 1.5 and 2 GeV/$c$. Leptons from $B \rightarrow D \ell \nu_{\ell}$ will usually have lower momenta than $B \rightarrow D^* \ell \nu_{\ell}$, because for $\vec{p}_D \rightarrow 0$, the  decay $B \rightarrow D \ell \nu_{\ell}$ is suppressed, so $D$ mesons will have higher momenta. Leptons from $B \rightarrow D^{**}(D^{(*)}\pi) \ell \nu_{\ell}$ will have much lower momenta, due to the reduced phase space available.

The multiplicity of reconstructed charged tracks in the event, in excess of those accounted for the $B_{tag}$ recoil side, the lepton and the $D^{0,+}$, offers as the missing mass a very high discriminating power to separate the different $B$ semileptonic decays. Every charged track is required to have its point of closest approach to the 
interaction point less than 10 cm along
the beam axis and less than 1.5 cm transverse to the beam axis. 
The number of additional tracks should be zero for pure $B \rightarrow D^{0,+} \ell \nu_{\ell}$ decays and larger than zero for 
those decays involving heavier charm states, provided these decay into a lighter $D$ accompanied by 
charged particles. 

\begin{figure}[!h]
\centering
\epsfig{figure=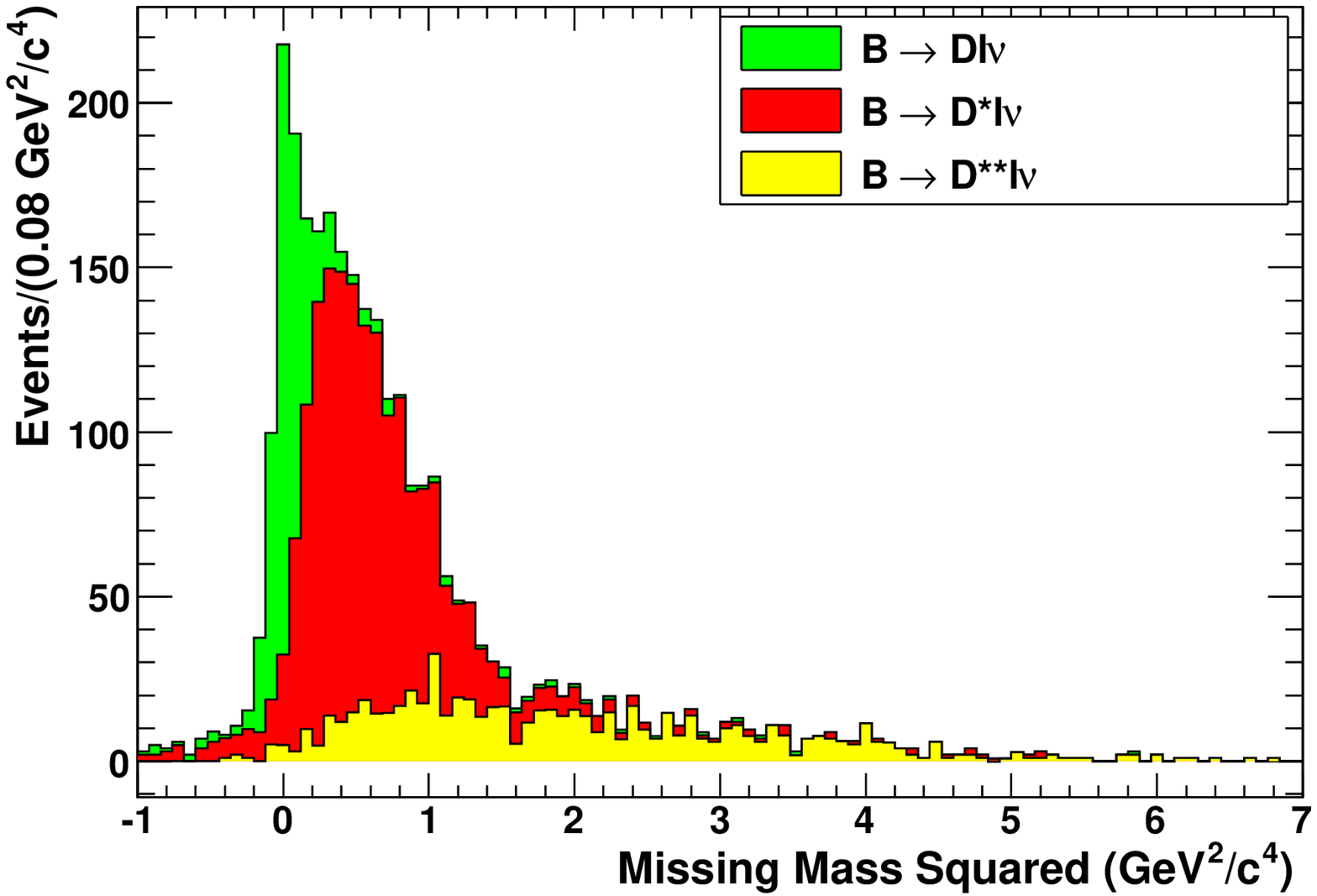,width=7.4cm}
\epsfig{figure=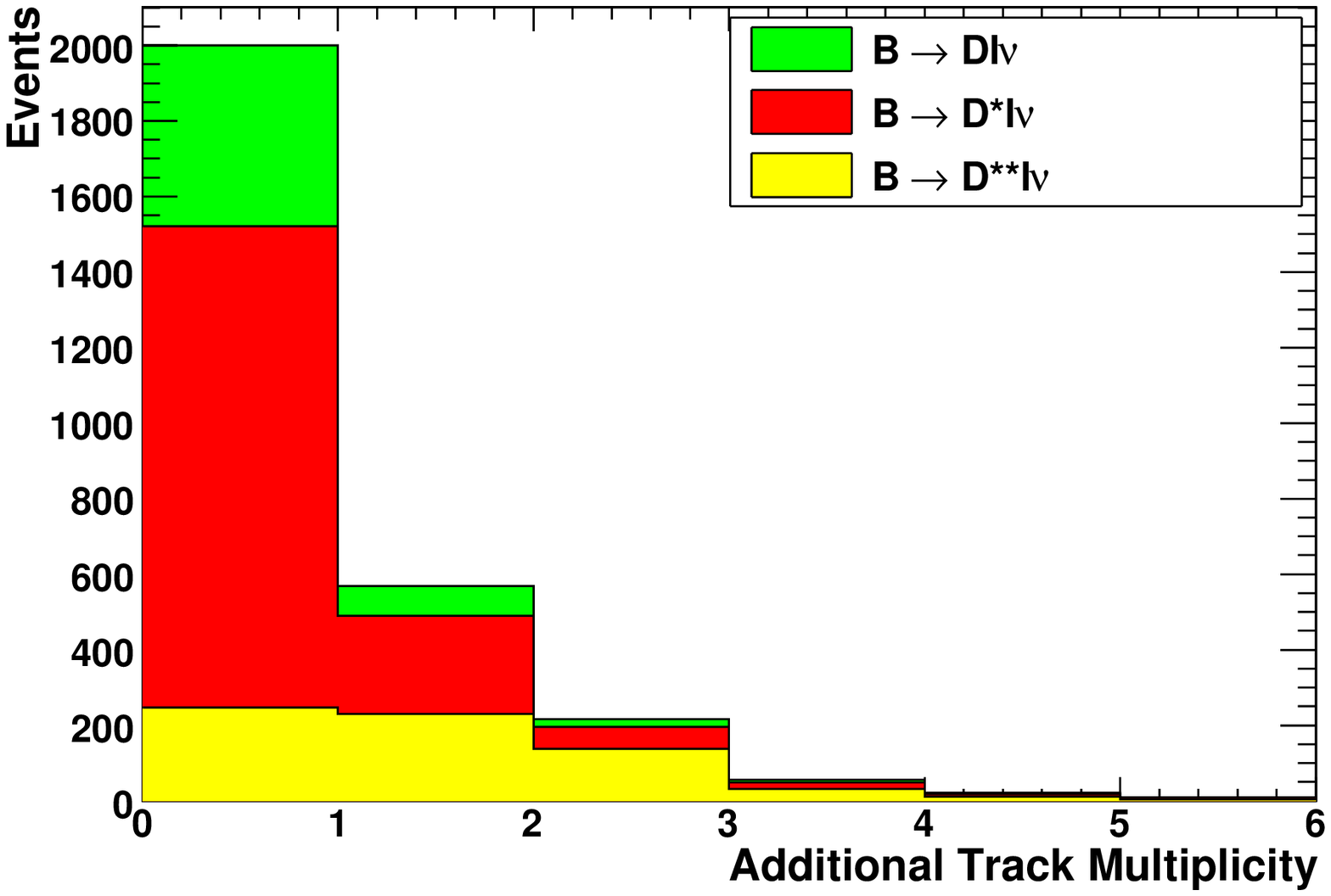,width=7.4cm}
\epsfig{figure=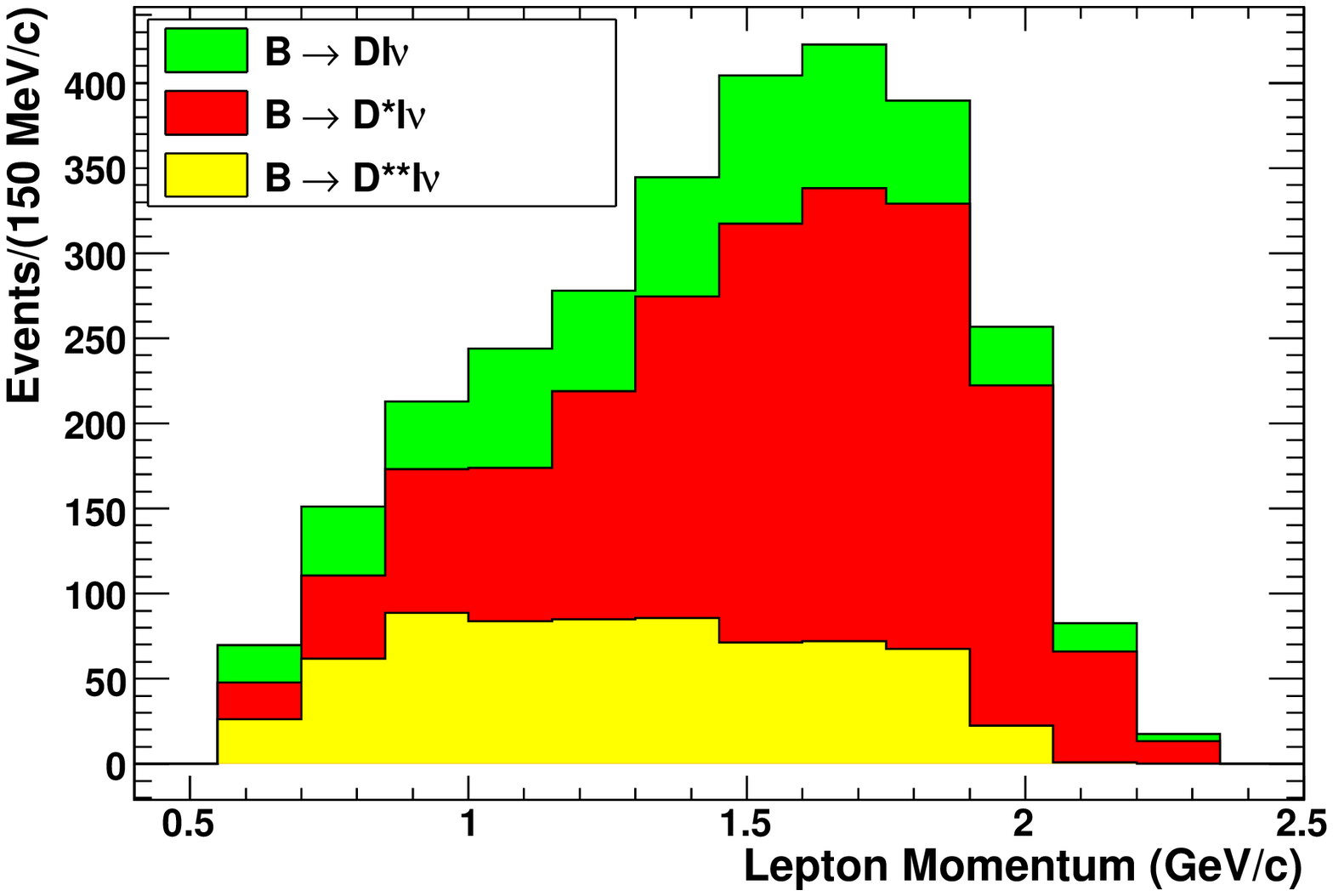,width=7.4cm}
\caption{Distributions of the missing mass squared, the multiplicity of additional charged tracks and the lepton momentum spectrum for a Monte Carlo sample of $B^- \rightarrow D X \ell^- \bar{\nu}_{\ell}$ events used for the  relative branching ratio $\Gamma(B^- \rightarrow D^{0,*0,**0} \ell^- \bar{\nu}_{\ell})/\Gamma (B^- \rightarrow D X \ell^- \bar{\nu}_{\ell})$ measurement; the different components are superimposed with different colors.}
\label{fig:cartoon}
\end{figure}

Figure \ref{fig:cartoon} shows the reconstructed missing mass squared, lepton momentum and additional charged track multiplicity distributions for a Monte Carlo sample of $B^- \rightarrow D X \ell^- \bar{\nu}_{\ell}$ events.  The different shapes of the  $B^- \rightarrow D^{(*,**)0} \ell^- \bar{\nu}_{\ell}$ components are clearly visible. As it can be observed, a fraction of  $B^- \rightarrow D^{(*)0} \ell^- \bar{\nu}_{\ell}$ events shows a multiplicity of additional charged tracks greater than zero, coming from badly reconstructed tracks not used for the $B_{tag}$ reconstruction. 

\section{Reconstruction of the Exclusive Decays $B^- \rightarrow D^{(*,**)0} \ell^- \bar{\nu}_{\ell}$}

To extract the relative branching ratios of the $B^- \rightarrow D^{(*,**)0} \ell^- \bar{\nu}_{\ell}$ decays from the inclusive $B^- \rightarrow D X \ell^- \bar{\nu}_{\ell}$ sample, Probability Density Functions (PDFs) corresponding to the different components present in the inclusive distributions must be built. These PDFs are determined using data sets enriched in the different exclusive $B^- \rightarrow D^{(*,**)0} \ell^- \bar{\nu}_{\ell}$ decays and then parameterized as analytical functions, with the exception of the charged particle track multiplicities which, owing to their discrete nature, are best described by histograms.

The selection of these enriched samples is optimized in terms of their purity, to introduce minimum biases with respect to the true underlying distributions. Background contributions in the exclusive samples include other $B$ semileptonic decays, where one particle is not reconstructed or erroneously added to the charm meson candidate (feed-down and feed-up). 
We use a sample of $B^- \rightarrow D^{0,+} X \ell^- \bar{\nu}_{\ell}$ or $B^- \rightarrow D^{*0,+} X \ell^- \bar{\nu}_{\ell}$ events to reconstruct the exclusive decays. $D^0$ and $D^+$ candidates are selected as above. $D^*$ mesons are reconstructed by combining a $D$ candidate and a pion in the $D^{*+} \rightarrow D^0 \pi^+ (D^+\pi^0)$ and $D^{*0} \rightarrow D^0 \pi^0$ decays. These $D^*$ candidates are required to have a mass difference $\Delta m = m_{D\pi} -m_D$ within $\pm 2,3,4$ MeV/$c^2$ of its nominal values for the decays $D^{*+} \rightarrow D^+ \pi^0$, $D^{*0} \rightarrow D^0 \pi^0$ and $D^{*+} \rightarrow D^0 \pi^+$ respectively. For the decays with a photon, $D^{*0} \rightarrow D^0 \gamma$, we require the mass difference $\Delta m = m_{D\gamma} -m_D$ within $\pm 9$ MeV/$c^2$ of its nominal value. $D^{**0}$ states or non resonant $B^- \rightarrow D^{(*)}\pi \ell^- \bar{\nu}_{\ell}$ decays are reconstructed by combining a $D$ or $D^{*}$ candidate with a charged pion. Neutral pions are not used at this stage of the reconstruction to avoid large combinatoric background. 

Using the four-momentum of the reconstructed $D^{*(**)}$ candidate, we compute the corresponding missing mass squared  $m^2_{miss,D^{(*,**)}}$ in the event, defined as:

\begin{equation}
 m^2_{miss,D^{(*,**)}} = (p(\Upsilon) -p(B_{tag}) - p(D^{*(**)}) - p(\ell))^2.
\end{equation}

 For true $B^- \rightarrow D^{(*,**)0} \ell^- \bar{\nu}_{\ell}$ semileptonic decays, this missing mass squared will  peak at zero. Exclusive samples are then selected by applying the relevant cuts on this variable to select or remove, if we want to eliminate feed-down components, a specific $B$ semileptonic decay.

To select 
$D^{*0} \ell^- \bar{\nu}_{\ell}$ decays, the variable $m^2_{miss, D^{*0}}$ is required to be in the range 
$-0.35 < m^2_{miss, D^{*0}} < 0.5$ GeV$^2/c^4$, while the unwanted decays are 
removed, by vetoing the peaking region in the corresponding missing mass squared distribution, i.e. $m^2_{miss, D^{**+}} > 0.5$ GeV$^2/c^4$ or $m^2_{miss, D^{**+}} < -0.6$ GeV$^2/c^4$ and 
$m^2_{miss, D^{**0}} < -0.2$ GeV$^2/c^4$. 

The exclusive decay $B^- \rightarrow D^0 \ell^- \bar{\nu}_{\ell}$ is selected by removing feed-down from $D^*$ and $D^{**}$ events. We require $m^2_{miss, D^{*0}} < -0.05$ GeV$^2/c^4$ to reduce feed-down from  $D^{*0} \rightarrow D^0 \gamma$ decays, $m^2_{miss, D^{*0}} < -0.4$ GeV$^2/c^4$ or $m^2_{miss, D^{*0}} > 1.0$ GeV$^2/c^4$ for $D^{*0} \rightarrow D^0 \pi^0$ decays. Feed-down from $D^{*+}$ events is reduced by requiring $|m^2_{miss, D^{*+}}| > 0.3$ GeV$^2/c^4$. Feed-down from $D^{**}$ events is reduced by requiring $|m^2_{miss, D^{**0}}| > 0.8$ GeV$^2/c^4$ or  $m^2_{miss, D^{**0}} < -0.5$ GeV$^2/c^4$ for $D^{**0} \rightarrow D^{*+} \pi^-$ and $D^{**0} \rightarrow D^{*0} \pi^0$ decays and $m^2_{miss, D^{**+}} < -0.5$ GeV$^2/c^4$ for $D^{**+}$ events. 

Exclusive $B^- \rightarrow D^{**0} \ell^- \bar{\nu}_{\ell}$ events are selected by requiring $-0.25 < m^2_{miss, D^{**+}} < 0.85 (0.25)$ GeV$^2/c^4$ for $D^{**0} \rightarrow D^+ (D^{*+}) \pi^-$ decays.  

Non $B$ semileptonic decay background and combinatoric \BB\ background are subtracted as in the inclusive distribution.
 Table \ref{tab:BPurity} lists the composition of the reconstructed exclusive samples in terms of purity,  feed-up(-down) fractions and non $B$ semileptonic background.

\begin{table}[!htb]
\centering
\caption{Purity and feed-down(up) contributions for the reconstructed exclusive decay modes.}
\begin{small}
\begin{tabular}{|c|c|c|c|c|c|c|}
\hline 
\hline
{\scriptsize Decay Mode}  &  {\scriptsize Purity}  & {\scriptsize Feed-up ($D$)} & {\scriptsize Feed-up ($D^*$)} & {\scriptsize Feed-down ($D^*$)} & {\scriptsize Feed-down ($D^{**}$)} &  {\scriptsize background}\\
\hline
{\scriptsize $B^- \rightarrow D^0 \ell^- \bar{\nu}_{\ell}$} & 75\% & - & - & 19.3\% & 3.6\% & 2.5\% \\
{\scriptsize $B^- \rightarrow D^{*0} \ell^- \bar{\nu}_{\ell}$} & 91\% & 3.7\% & - & - & 2.8\% & 2.6\%\\
{\scriptsize $B^- \rightarrow D^{**0} (D^{*+} \pi^-) \ell^- \bar{\nu}_{\ell}$} & 85\% & 0.27\% & 5.5\% & - & - & 9.4\%\\
{\scriptsize $B^- \rightarrow D^{**0} (D^+ \pi^-)\ell^- \bar{\nu}_{\ell}$} & 85\% & 1.2\% & 4.2\% & - & - & 9.7\%\\
\hline
\hline
\end{tabular}
\end{small}
\label{tab:BPurity}
\end{table}

This reconstruction strategy has been validated to assess whether the exclusive
selection procedure induces biases on the shapes of the discriminating variables.
This has been done by comparing the distribution of these variables using different selection criteria on Monte Carlo simulated events of the different decay modes $B^- \rightarrow D^{(*,**)0} \ell^- \bar{\nu}_{\ell}$. 
The agreement of these distributions is good: residual discrepancies are evaluated as systematic effects.

The final step is the determination of the fractions
of $D^0$, $D^{*0}$ and $D^{**0}$ decays in the selected inclusive semileptonic
sample, by a simultaneous $\chi^2$ fit to the missing mass squared, additional charged track multiplicity and lepton momentum distributions 
where the relative fractions are
treated as free parameters together with the parameters describing the shapes of
the discriminating variables. The PDFs for the three different fit components are extracted by fitting simultaneously the corresponding exclusive and inclusive reconstructed distributions.  
The $D^0$ missing mass squared PDF is  the sum of three Gaussians, with independent means and widths, to account for the upper tail, which is mainly due to feed-down from $B^- \rightarrow D^{*0} \ell^- \bar{\nu}_{\ell}$ decays where the soft pion or the photon from the $D^{*0}$ is missing. The $D^{*0}$ ($D^{**0}$) missing mass squared PDF is built by the product of a Gaussian (an exponential) with a polynomial function. 
The three $D^0$, $D^{*0}$ and $D^{**0}$ lepton momentum spectra are parameterized by the product of a polynomial and an exponential which describes the low  momentum tail of the spectra. 
Because we do not reconstruct $D^{**0}$ states involving neutral pions, we use the $B^- \rightarrow D^{*0} \ell^- \bar{\nu}_{\ell}$ additional charged track multiplicity distribution to model $B^- \rightarrow D^{**0} \ell^- \bar{\nu}_{\ell}$ events, where the $D^{**0}$ decays in neutral pions.

The total PDF used to model the inclusive distribution is obtained by the sum of the three exclusive PDFs, whose relative weights correspond to the relative fractions of $D^0$, $D^{*0}$ and $D^{**0}$ decays in the sample. The sum of the relative weights is constrained to be equal to one.
This results in a $33$-parameter fit which ensures
that statistical correlations are properly taken into account and the
uncertainties on the exclusive shapes, obtained on much reduced statistics
compared to the inclusive sample, are correctly propagated into the
statistical uncertainties on the $D^0$, $D^{*0}$ and $D^{**0}$ fractions. The fit 
also takes into account the feed-up and feed-down fractions in the exclusive shapes, 
obtained from simulation and contributing to the systematic uncertainty.

\begin{figure}
\centering
\epsfig{figure=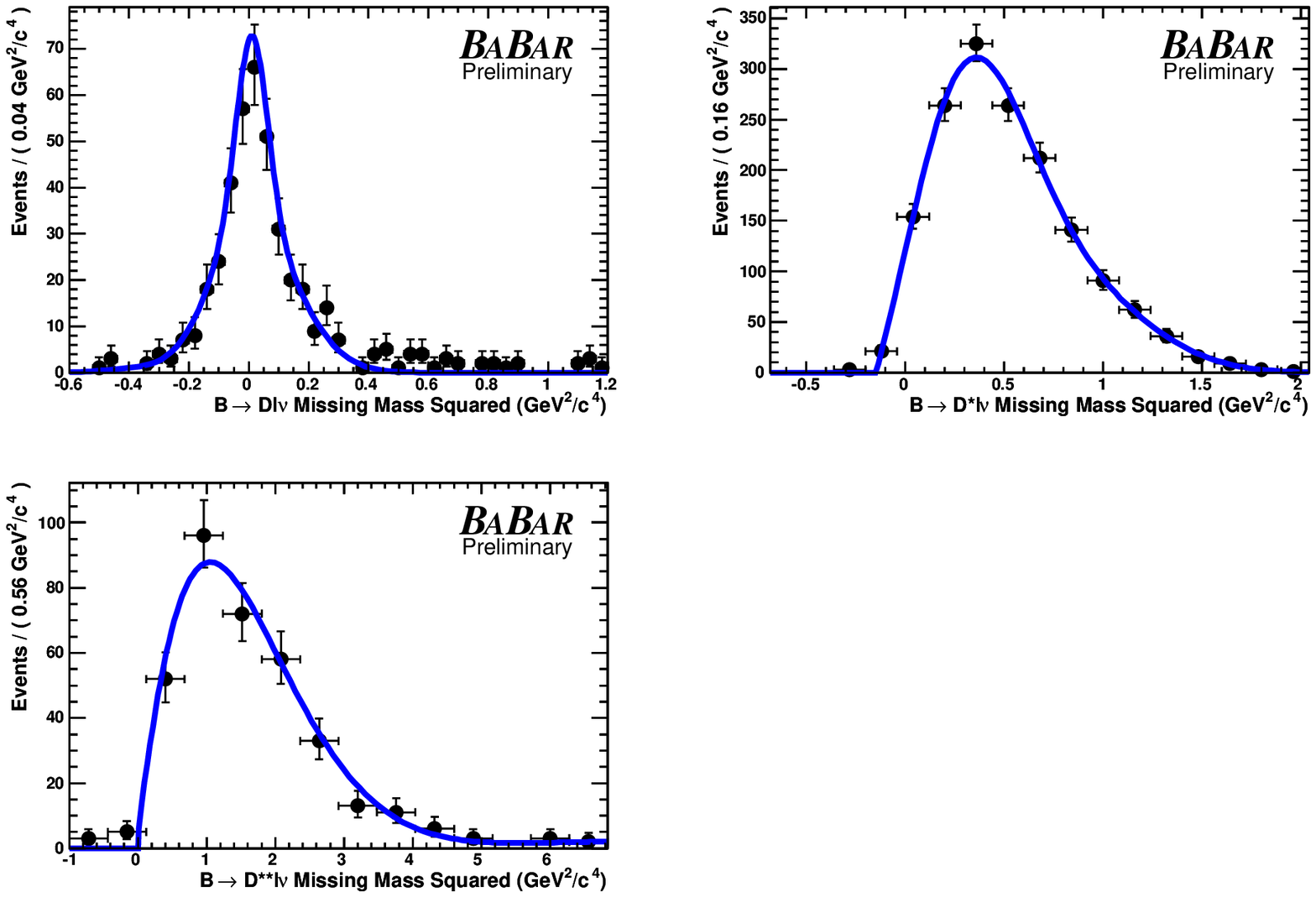,width=14.8cm}
\epsfig{figure=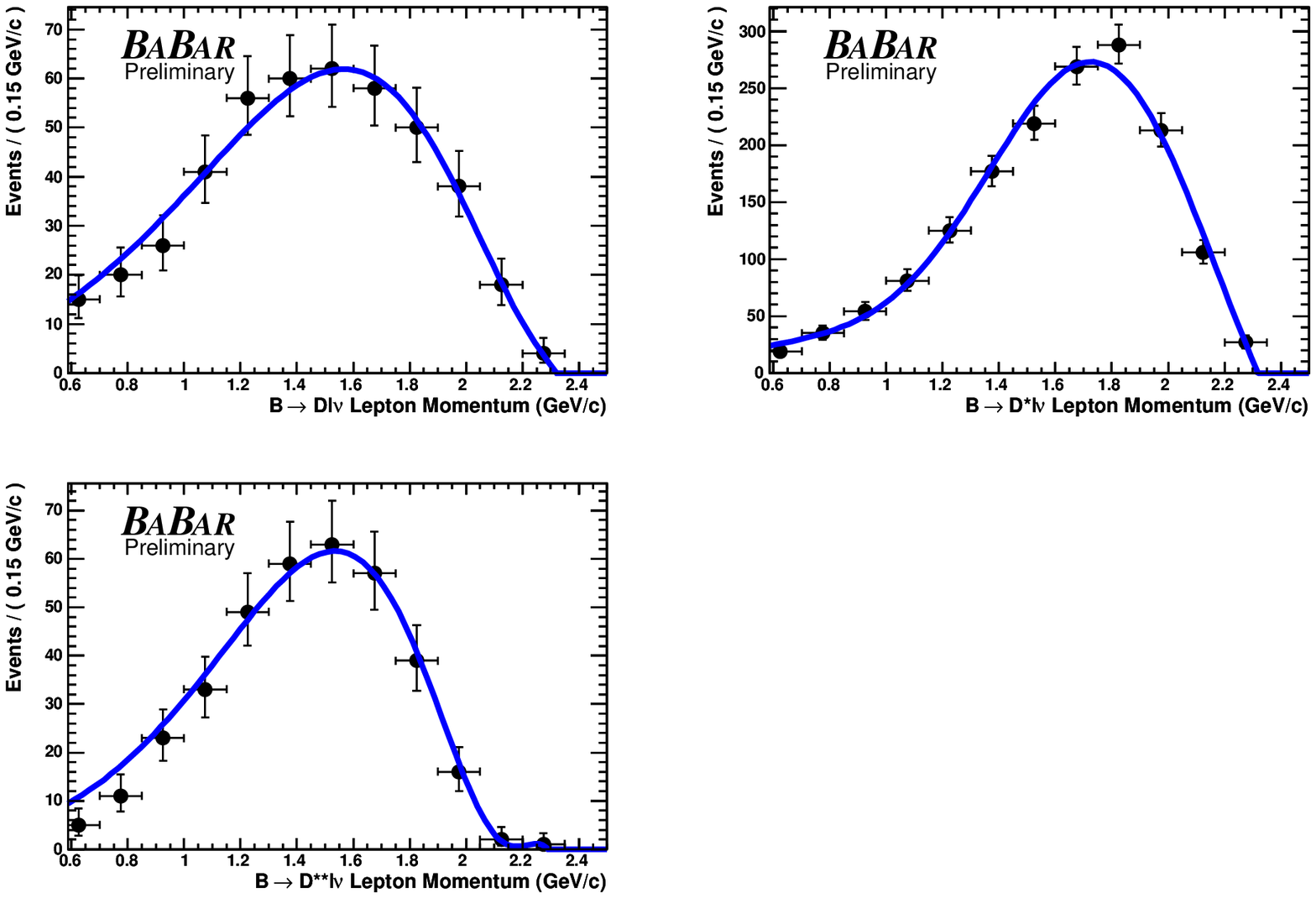,width=14.8cm}
\caption{Missing mass squared and lepton momentum distributions for the three selected exclusive decay modes, $B^- \rightarrow D^0 \ell^- \bar{\nu}_{\ell}$, $B^- \rightarrow D^{*0} \ell^- \bar{\nu}_{\ell}$ and $B^- \rightarrow D^{**0} \ell^- \bar{\nu}_{\ell}$. The data are compared to the result of the overall fit (solid line) to the distributions for the selected exclusive and inclusive samples.}
\label{fig:FitRes1}
\end{figure}

\begin{figure}
\centering
\epsfig{figure=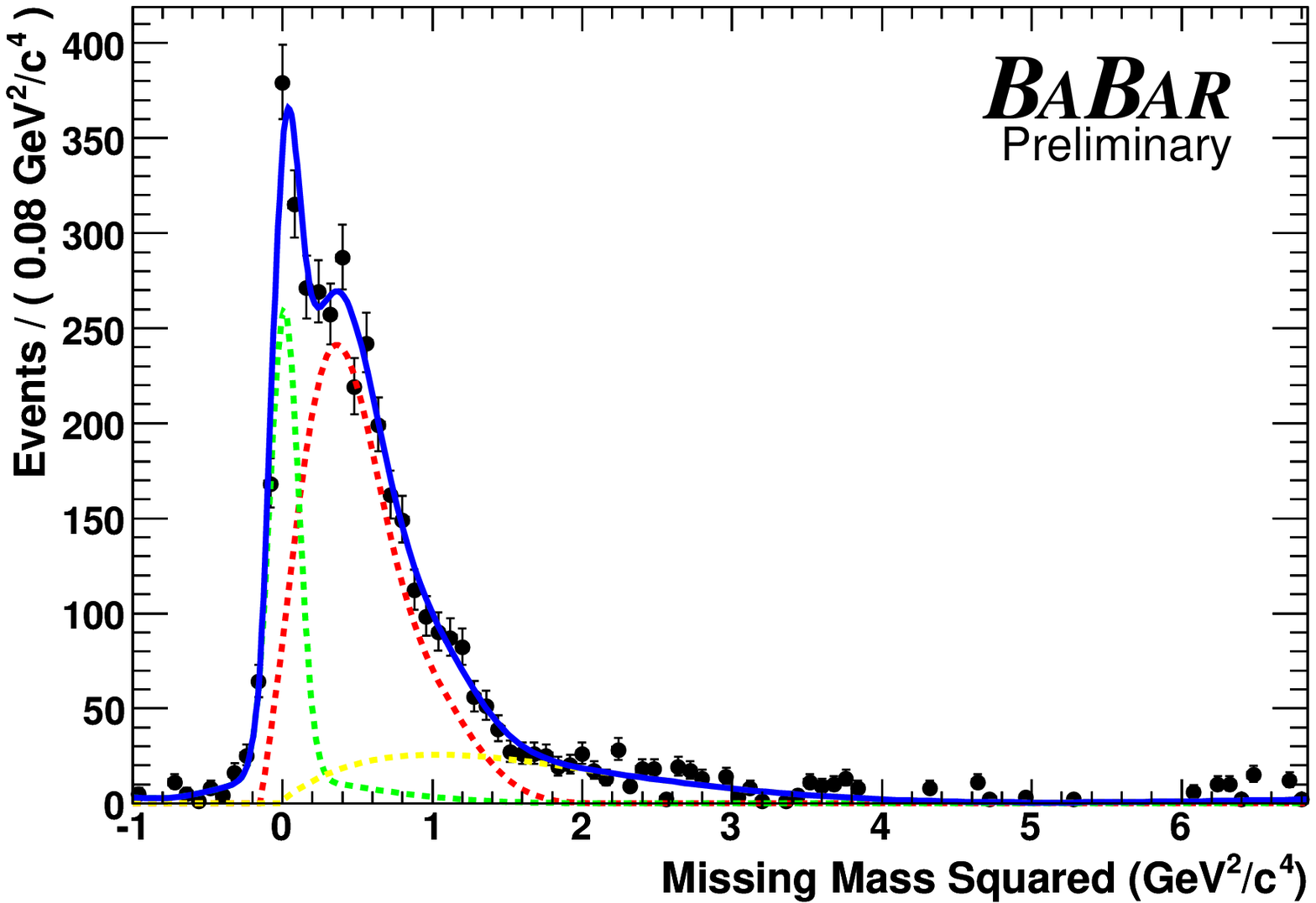,width=7.4cm}
\epsfig{figure=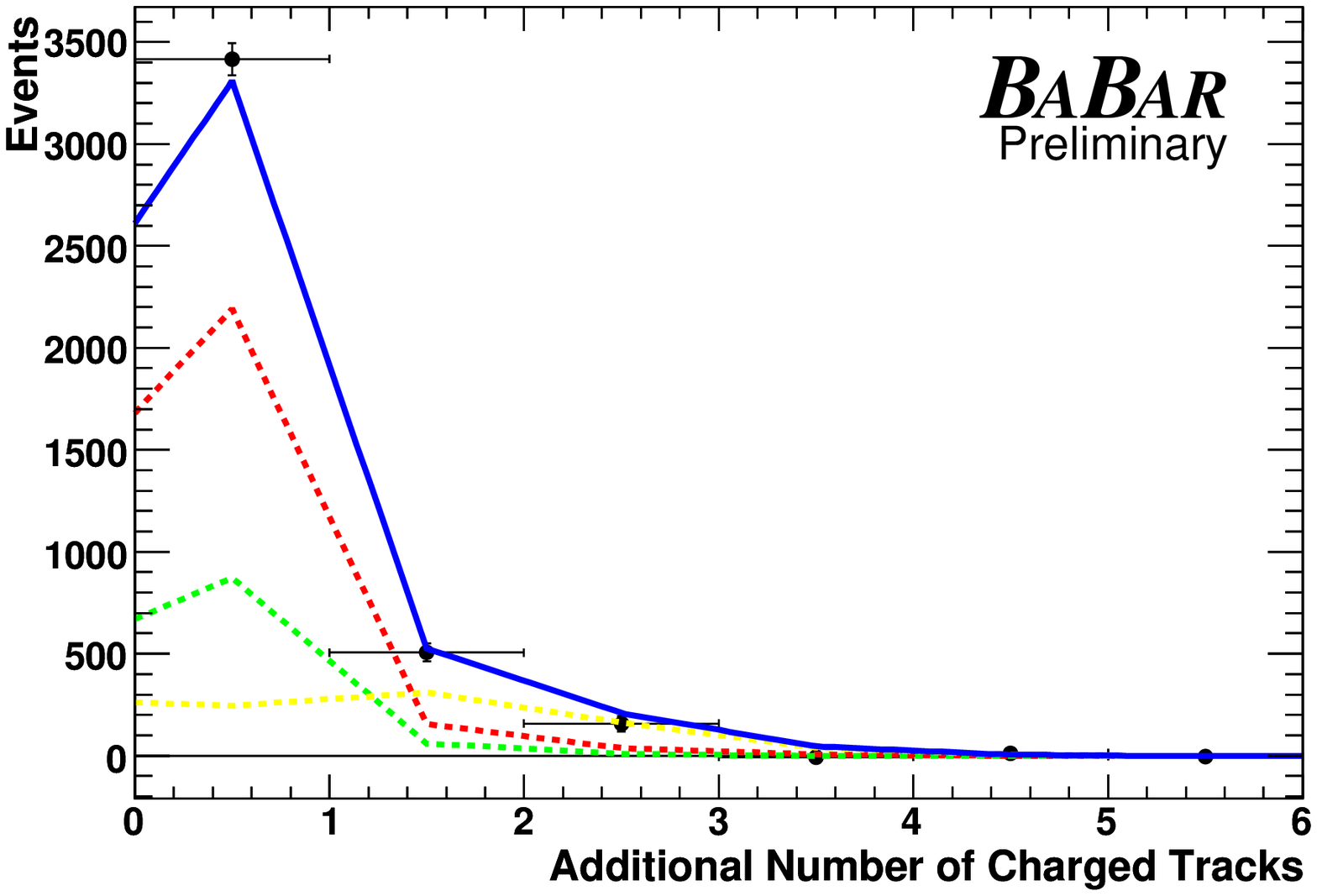,width=7.4cm}
\epsfig{figure=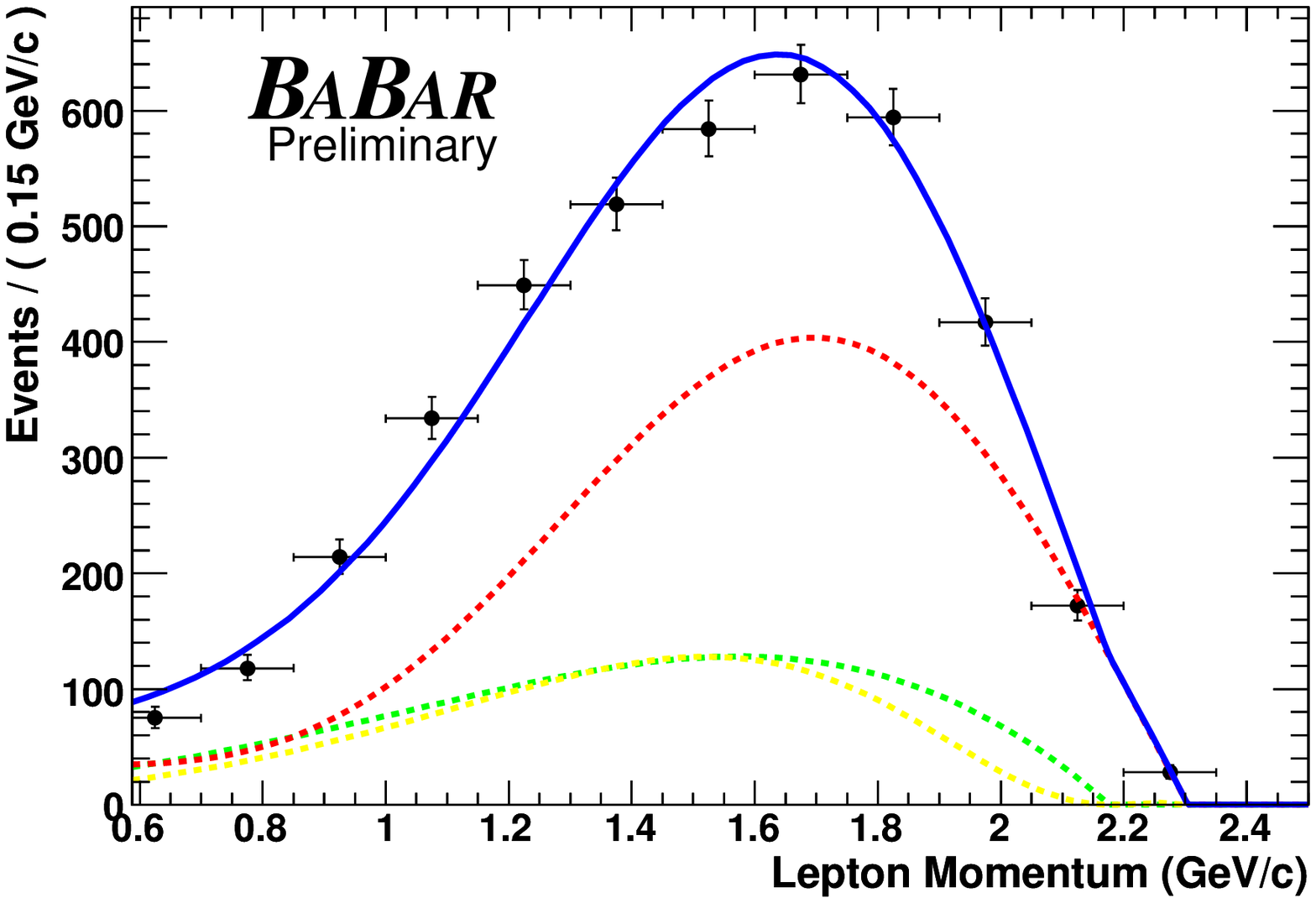,width=7.4cm}
\caption{Distributions of the missing mass squared, the multiplicity of additional charged tracks and the lepton momentum spectrum for the selected $B^- \rightarrow D X \ell^- \bar{\nu}_{\ell}$ events. The data are compared to the results of the overall fit (solid line). The PDFs for the three $D^{(*,**)0}$ components are shown in different colors, from lighter ($D^{**0}$) to darker ($D^{*0}$).
}
\label{fig:FitRes2}
\end{figure}

The fit results for the $B^- \rightarrow D X \ell^- \bar{\nu}_{\ell}$ are shown in Figure \ref{fig:FitRes1} and \ref{fig:FitRes2}. Figure \ref{fig:FitRes1} shows the fit result for the exclusive missing mass squared and lepton momentum distributions, which are used to build the PDF used to fit the corresponding inclusive distributions, shown in 
Figure \ref{fig:FitRes2}. The simultaneous fit has a total $\chi^2$/ndof = 237/205.

\section{Systematic Studies}
\label{sec:Systematics}

Different sources of systematic uncertainties have been estimated. These are divided in various 
categories. Detector related systematics may arise from 
discrepancies in the simulation modelling of the track reconstruction and efficiency, particle 
identification and neutral particle reconstruction. 

 Uncertainties related to the reconstruction of charged tracks are determined by evaluating the fit stability using different track selection criteria and by a toy Monte Carlo study in which we fluctuate the track multiplicity bin content according to the tracking efficiency uncertainty.
The systematic error due to the reconstruction of neutral particles is studied by varying the resolution and efficiency to match those found in control samples on data. 
We estimate the systematic error due to particle identification by varying the electron and muon identification efficiencies by 2\% and 3\%, respectively. The misidentification probabilities are varied by 15\% for both electrons and muons.

Different systematic uncertainties are evaluated for the inclusive and exclusive reconstruction procedure. 
For the inclusive reconstruction of the $B^- \rightarrow D X \ell^- \bar{\nu}_{\ell}$ samples, the weighting factors used to subtract $B$ cascade decays are varied within their errors and the differences in the fitted relative branching ratios are included in the systematic uncertainty. We also evaluate systematic effects associated to the  misidentification rate and the tracking efficiency of the photon conversion and $\pi^0$ Dalitz decay reconstruction algorithm.
The uncertainty of the $B_{tag}$ combinatorial background subtraction is estimated by evaluating differences in the shapes of this background in the sideband and in the signal region using Monte Carlo predictions. We also evaluate the dependence of the sideband scaling factors on the lepton momentum.
The systematic error due to the uncertainty in the amount of flavor cross-feed is computed by varying its fraction  by a conservative 30\%. 
The effect of uncertainty on reconstruction efficiency mostly cancels in the relative branching ratios; the uncertainty due to the finite Monte Carlo statistics and the $m_{ES}$ fit we use to compute the relative reconstruction efficiency is included accordingly in the systematic error evaluation.

For the exclusive reconstruction of $B^- \rightarrow D^{(*,**)0} \ell^- \bar{\nu}_{\ell}$ decays, systematic uncertainties arise from the evaluation of the purity of each sample and residual biases on the reconstructed 
distributions due to the selection criteria. 
Possible mismatching between data and Monte Carlo is accounted for by varying the feed-down and feed-up
fractions within their errors and by evaluating as systematic uncertainties differences in the fit results.
We also evaluate possible biases due to the use of reconstructed exclusive
distributions to build the PDFs by replacing them with Monte Carlo signal
distributions and performing the fit using these shapes. A separate systematic uncertainty is computed for the $D^{**0}$ additional track multiplicity distribution by varying the relative ratio of $D^{**0} \rightarrow D^{(*)+} \pi^-$ and $D^{**0} \rightarrow D^{(*)0} \pi^0$ events which is predicted by our Monte Carlo simulation. 
The uncertainty of the $B_{tag}$ combinatorial background subtraction and the cross-feed correction is computed as for the inclusive reconstruction.

The fit performance has been extensively tested using simulated samples with
varying fractions of the different decay modes. These tests show that the
procedure adopted in this analysis is able to extract the decay fractions
without significant biases;  the statistical errors obtained from the fit are corrected according to the results on independent toy Monte Carlo tests and increased by 15(20)\% for the $D^{(*)0}(D^{**0})$ component.
We evaluated the systematics due to the fit technique by using
different parameterizations of the exclusive shapes. Additional
cross-checks were made, repeating the simultaneous fit of the exclusive
shape parameters and the exclusive yields in different configurations,
e.g. by fixing the PDF parameters for the exclusive distributions or by
only fitting the inclusive distributions.

\section{Results}
\label{sec:Physics}

The extraction of the relative fractions of $D^0$, $D^{*0}$ and $D^{**0}$ decays in the selected inclusive semileptonic
sample gives the relative branching ratios $\Gamma(B^- \rightarrow D^{(*,**)0} \ell^- \bar{\nu}_{\ell})
 /\Gamma(B^- \rightarrow D X \ell^- \bar{\nu}_{\ell})$. 

The measured branching ratios are reported in Table \ref{tab:fitResBplusFit}  
and the different systematic uncertainties are shown in Table \ref{tb:systB}.  

\begin{table}[!htb]
\centering
\caption{Fitted branching ratios $\Gamma(B^- \rightarrow D^{0,*0,**0} \ell^- \bar{\nu}_{\ell})/\Gamma (B^- \rightarrow D X \ell^- \bar{\nu}_{\ell})$ with the corresponding statistical uncertainty.}
\begin{tabular}{|c|c|c|}
\hline
\hline
Branching Ratio   & Fit Results \\
\hline
$\frac{\Gamma(B^- \rightarrow D^0 \ell^- \bar{\nu}_{\ell})}{\Gamma (B^- \rightarrow D X \ell^- \bar{\nu}_{\ell})}$  &  (21.0 $\pm$ 1.44) \%\\ 
\hline
$\frac{\Gamma(B^- \rightarrow D^{*0} \ell^- \bar{\nu}_{\ell})}{\Gamma (B^- \rightarrow D X \ell^- \bar{\nu}_{\ell})}$  & (61.1 $\pm$ 1.94)\%\\
\hline
$\frac{\Gamma(B^- \rightarrow D^{**0} \ell^- \bar{\nu}_{\ell})}{\Gamma (B^- \rightarrow D X \ell^- \bar{\nu}_{\ell})}$ & (17.3 $\pm$ 1.44) \%\\
\hline
\hline
\end{tabular}
\label{tab:fitResBplusFit}
\end{table}

\begin{table}[!htb]
\centering
\caption{Systematic uncertainties in the measurement of $\Gamma(B^- \rightarrow D^{(*,**)0} \ell^- \bar{\nu}_{\ell})/\Gamma(B^- \rightarrow D X \ell^- \bar{\nu}_{\ell})$.}
\begin{tabular}{|c|c|c|c|}
\hline
\hline
 &\multicolumn{3}{c|}{{\scriptsize Systematic uncertainty on $\Gamma(B^- \rightarrow D^{(*,**)0}\ell^- \bar{\nu}_{\ell}) /\Gamma(B^- \rightarrow D X \ell^- \bar{\nu}_{\ell})$} } \\
\hline
& $B^- \rightarrow D^0\ell\nu_{\ell}$ & $B^- \rightarrow D^{*0}\ell\nu_{\ell}$ & $B^- \rightarrow D^{**0}\ell\nu_{\ell}$ \\
\hline
Tracking efficiency & 0.009  & 0.008 & 0.004\\
Neutral reconstruction & 0.001 & 0.003 & 0.0009\\
Electron ID & 0.0008  & 0.002 & 0.0007\\
Muon ID & 0.006  & 0.02 & 0.005\\
\hline
Inclusive Reconstruction & \multicolumn{3}{c|}{} \\
\hline
Cascade decay background & 0.01 & 0.01 & 0.01\\
Conversion and Dalitz decay background & 0.001  & 0.004 & 0.001\\
MC statistics and $M_{ES}$ fit & 0.012  & 0.01 & 0.011\\
Cross-feed corrections & 0.001  & 0.002 & 0.004\\
\hline
Exclusive Reconstruction & \multicolumn{3}{c|}{}\\
\hline
Feed-down and feed-up corrections & 0.003  & 0.002 & 0.002\\
MC signal shapes & 0.002 & 0.002 & 0.007\\
$D^{**0}$ track multiplicity & 0.005  & 0.002 & 0.005\\
$M_{ES}$ fit & 0.0007  & 0.004 & 0.003\\
Cross-feed corrections & 0.0003  & 0.0006 & 0.0009\\
\hline
Fit Technique & \multicolumn{3}{c|}{}\\
\hline
Missing mass and lepton momentum PDF & 0.007 & 0.002 & 0.008\\
\hline
Total Systematic Error & 0.021 & 0.027 & 0.021 \\
\hline
\hline
\end{tabular}
\label{tb:systB}
\end{table}

No direct measurement of the relative branching ratios  $\Gamma(B^- \rightarrow D^{(*,**)0} \ell^- \bar{\nu}_{\ell})
 /\Gamma(B^- \rightarrow D X \ell^- \bar{\nu}_{\ell})$ is present in the literature. 
In the absence of any measurement~\cite{pdg} for $B$ semileptonic decays into charm final states $X_c$ other than $D$ meson states, we can assume the relation: 

\begin{equation}
\frac{\Gamma(B^- \rightarrow D^{(*,**)0} \ell^- \bar{\nu}_{\ell})}
 {\Gamma(B^- \rightarrow X_c \ell^- \bar{\nu}_{\ell})}=\frac{\Gamma(B^- \rightarrow D^{(*,**)0} \ell^- \bar{\nu}_{\ell})}
 {\Gamma(B^- \rightarrow D X \ell^- \bar{\nu}_{\ell})}
\end{equation}

\noindent which can be used with the current best values~\cite{hfag} for $B^- \rightarrow X \ell \nu_{\ell}$, $b \rightarrow u \ell \nu_{\ell}$ and the branching fractions $B^- \rightarrow D^{(*,**)0} \ell \bar{\nu}_{\ell}$ to make a comparison with our results.

The $B^- \rightarrow D^{**0}\ell^-\bar{\nu}_{\ell}$ component is also sensitive to the presence of $B^- \rightarrow D^{(*)}n\pi\ell^-\bar{\nu}_{\ell}$
events in the inclusive sample $B^- \rightarrow D X\ell^-\bar{\nu}_{\ell}$, despite we do not explicitly
introduce in the fit the corresponding PDF. By floating its parameters in the
fit, the D** PDF can model any excess of events in the inclusive sample $B^- \rightarrow D X\ell^-\bar{\nu}_{\ell}$, which can be associated to a small $B^- \rightarrow D^{(*)}n\pi\ell^-\bar{\nu}_{\ell}$ component.

\section{Conclusions}
\label{sec:Summary}

Preliminary results of the relative branching fractions of individual semileptonic $B$ decays have been obtained for the $B^-$ decaying into $D^0 \ell^- \bar{\nu}_{\ell}$, $D^{*0} \ell^- \bar{\nu}_{\ell}$ and $D^{**0} \ell^- \bar{\nu}_{\ell}$. These measurements are based on a sample of 239 million $\Upsilon(4S) \rightarrow B\bar{B}$ decays collected by the \babar\ experiment at the \pep2\ asymmetric-energy $B$ factory at SLAC, in events in which one $B$ meson decaying to a hadronic final state is fully reconstructed.  
A novel technique, based on a multi-parametric fit to a set of discriminating variables in an inclusive sample of $B^- \rightarrow D X \ell^- \bar{\nu}_{\ell}$ events, has been used to measure
the relative branching fractions for $B^- \rightarrow D^0, D^{*0}$ and $D^{**0}(D^{(*)} \pi) \ell^- \bar{\nu}_{\ell}$ 
decays.
In order to reduce the sensitivity to Monte Carlo modelling, which is especially limited by the poor knowledge of the $B^- \rightarrow D^{**0} (D^{(*)} \pi) \ell^- \bar{\nu}_{\ell}$ decays, the shapes of the discriminating
variables for the $D^0$, $D^{*0}$ and $D^{**0}$ decays are extracted from selected data samples of exclusive decays.

Correcting the statistical errors as discussed in section 5, we obtain:

\begin{eqnarray}
\frac{\Gamma(B^- \rightarrow D^0 \ell^- \bar{\nu}_{\ell})} {\Gamma(B^- \rightarrow D X \ell^- \bar{\nu}_{\ell})} &=&  0.210 \pm 0.017~(\mbox{stat.})  \pm  0.021~(\mbox{syst.})  \nonumber \\ 
\frac{\Gamma(B^- \rightarrow D^{*0} \ell^- \bar{\nu}_{\ell})} {\Gamma(B^- \rightarrow D X \ell^- \bar{\nu}_{\ell})} &=&  0.611 \pm  0.022~(\mbox{stat.}) \pm 0.027~(\mbox{syst.})  \nonumber \\ 
\frac{\Gamma(B^- \rightarrow D^{**0} \ell^- \bar{\nu}_{\ell})} {\Gamma(B^- \rightarrow D X \ell^- \bar{\nu}_{\ell})} &=&  0.173 \pm 0.017~(\mbox{stat.}) \pm 0.021~(\mbox{syst.})  \nonumber \\ \nonumber
\end{eqnarray}

The BELLE result~\cite{Abe:2005up} is consistent with our measurement. 
The extension of this technique to the 
$\bar{B}^0 \rightarrow D^{(*,**)0} \ell^- \bar{\nu}_{\ell}$ decay modes  
 will increase the accuracy of the 
branching fraction measurements of $B$ semileptonic decay in $D$ meson final states 
and will help the understanding of the observed difference between the 
rate for the inclusive semileptonic decay
and the sum of the rates for the inclusive modes~\cite{pdg}. 
Moreover, an accurate measurement of the $B \rightarrow D^{**}\ell\nu$ branching 
fraction, which is made possible by the use of this method, 
is important to understand the decay mechanism of this channel, which provides
crucial tests of the Heavy Quark Effective Theory and the QCD Sum
rules~\cite{yaounanc}.

\section{Acknowledgments}
\label{sec:Acknowledgments}

We are grateful for the 
extraordinary contributions of our \pep2\ colleagues in
achieving the excellent luminosity and machine conditions
that have made this work possible.
The success of this project also relies critically on the 
expertise and dedication of the computing organizations that 
support \babar.
The collaborating institutions wish to thank 
SLAC for its support and the kind hospitality extended to them. 
This work is supported by the
US Department of Energy
and National Science Foundation, the
Natural Sciences and Engineering Research Council (Canada),
Institute of High Energy Physics (China), the
Commissariat \`a l'Energie Atomique and
Institut National de Physique Nucl\'eaire et de Physique des Particules
(France), the
Bundesministerium f\"ur Bildung und Forschung and
Deutsche Forschungsgemeinschaft
(Germany), the
Istituto Nazionale di Fisica Nucleare (Italy),
the Foundation for Fundamental Research on Matter (The Netherlands),
the Research Council of Norway, the
Ministry of Science and Technology of the Russian Federation, and the
Particle Physics and Astronomy Research Council (United Kingdom). 
Individuals have received support from 
the Marie-Curie IEF program (European Union) and
the A. P. Sloan Foundation.


\begin{thebibliography}{99}

\bibitem{Morenas:1997nk}
  V.~Morenas, A.~Le Yaouanc, L.~Oliver, O.~Pene and J.~C.~Raynal,
  Phys.\ Rev.\ D {\bf 56}, 5668 (1997)
  [arXiv:hep-ph/9706265].

\bibitem{Leibovich:1997em}
  A.~K.~Leibovich, Z.~Ligeti, I.~W.~Stewart and M.~B.~Wise,
  Phys.\ Rev.\ D {\bf 57}, 308 (1998)
  [arXiv:hep-ph/9705467].

\bibitem{Ebert:2000bj}
  D.~Ebert, R.~N.~Faustov and V.~O.~Galkin,
  arXiv:hep-ph/0006187.

\bibitem{DiPierro:2002eu}
  M.~Di Pierro and A.~K.~Leibovich,
  arXiv:hep-ph/0207134.

\bibitem{pdg}
S.~Eidelman {\it et al.}  [Particle Data Group],
Phys.\ Lett.\ B {\bf 592} 1 (2004).

\bibitem{Buskulic:1996uk}
  D.~Buskulic {\it et al.}  [ALEPH Collaboration],
  Z.\ Phys.\ C {\bf 73}, 601 (1997).

\bibitem{Abreu:2000}
  P.~Abreu {\it et al.}  [DELPHI Collaboration],
  Phys.\ Lett.\ B {\bf 475}, 407 (2000).

\bibitem{Bloch:2000}
  D.~Bloch {\it et al.}  [DELPHI Collaboration],
DELPHI~2000-106~CONF~405.

\bibitem{Abbiendi:2002ge}
  G.~Abbiendi {\it et al.}  [OPAL Collaboration],
  Eur.\ Phys.\ J.\ C {\bf 30}, 467 (2003)
  [arXiv:hep-ex/0301018].

\bibitem{Anastassov:1997im}
  A.~Anastassov {\it et al.}  [CLEO Collaboration],
  Phys.\ Rev.\ Lett.\  {\bf 80}, 4127 (1998)
  [arXiv:hep-ex/9708035].

\bibitem{Abazov:2005ga}
  V.~M.~Abazov {\it et al.}  [D0 Collaboration],
  Phys.\ Rev.\ Lett.\  {\bf 95}, 171803 (2005)
  [arXiv:hep-ex/0507046].

\bibitem{hfag}
Heavy Flavor Averaging Group,  
http://www.slac.stanford.edu/xorg/hfag/semi/winter06/winter06.shtml

\bibitem{Abe:2005up}
D.~Liventsev {\it et al.} [Belle Collaboration], Phys.\ Rev.\ D{\bf 72}, 051109 (2005).

\bibitem{ref:babar}
B.\ Aubert {\em et al.} [\babar\ Collaboration],
Nucl.\ Instrum.\ Methods A {\bf 479}, 1 (2002).

\bibitem{geant}
S.~Agostinelli {\em et al.} [GEANT 4 Collaboration],
Nucl.\ Instrum.\ Methods A {\bf 506}, 250 (2003).

\bibitem{CrystallBall}
J.~E.~Gaiser {\it et al.}, Phys.\ Rev.\ D {\bf 34}, 711 (1986). 

\bibitem{argusfunc}
H.~Albrecht {\it et al.} (ARGUS Collab.), Z.~Phys.~C {\bf 48}, 543 (1990). 

\bibitem{exampArgus}
``Exclusive semileptonic decays of B mesons to D mesons''
By ARGUS Collaboration (H. Albrecht et al.). DESY-92-029, Feb 1992.

\bibitem{yaounanc}
A.~Le~Yaounanc et al., Phys.~Lett.~B {\bf 520}, 25 (2001).

\end{thebibliography}
\end{document}